\documentstyle[12pt,aasms]{article}
\begin{document}
\title{Gravitational Radiation, Inspiraling Binaries, and Cosmology}
\author{David F.~Chernoff}
\affil{Astronomy Department, Cornell University, Ithaca, NY 14853}
\author{Lee Samuel Finn}
\affil{Department of Physics and Astronomy, Northwestern
University, Evanston, Illinois 60208}

\def\newacronym#1#2#3{\def#1{#3 (#2)\def#1{#2}}}

\newacronym{\ligo}{LIGO}{Laser Interferometer Gravitational-wave
Observatory}
\newacronym{\lli}{LLI}{LIGO-like Interferometer}
\newacronym{\ld}{LD}{LIGO-detector}

\newcommand{\scrA}[0]{{\cal A}}
\newcommand{\scrM}[0]{{\cal M}}
\newcommand{\scrN}[0]{{\cal N}}
\newcommand{\htns}{\hat{\mbox{\boldmath$n$}}_S}
\newcommand{\htnb}{\hat{\mbox{\boldmath$n$}}_B}
\newcommand{\chirpM}{{\scrM}}
\newcommand{\amp}{{\scrA}}
\newcommand{\cosmo}{\mbox{\boldmath $\mu$}}
\newcommand{\bzeta}{\mbox{\boldmath $\zeta$}}
\newcommand{\widcosmo}{\mbox{$\widetilde {\mu_i}$}}
\newcommand{\hatcosmo}{\mbox{$\hat {\mu_i}$}}
\newcommand{\delcosmo}{\mbox{$\delta {\mu_i}$}}
\newcommand{\twidrho}{\mbox{$\tilde\rho$}}
\newcommand{\kms}{\hbox{\rm km/s}}
\newcommand{\secs}{\hbox{\rm s}}
\newcommand{\km}{\hbox{\rm km}}
\newcommand{\Mpc}{\hbox{\rm Mpc}}
\newcommand{\Gpc}{\hbox{\rm Gpc}}
\newcommand{\msun}{\mbox{$M_\odot$}}
\newcommand{\yr}{\hbox{\rm yr}}
\newcommand{\Ora}{\hbox{\rm O}}

\newcommand{\Nt}{{\scrN}}
\newcommand{\Ntz}{{\scrN_{0}}}
\newcommand{\Ntzh}{\widehat{\scrN_{0}}}
\newcommand{\ndot}{\mbox{$\dot n$}}

\begin{abstract}
We show how to measure cosmological parameters
using observations of inspiraling binary neutron star or
black hole systems in one or more gravitational wave detectors.  To
illustrate, we focus on the case of fixed mass binary
systems observed in a single Laser Interferometer Gravitational-wave
Observatory (LIGO)-like detector.  Using realistic detector noise
estimates, we characterize the rate of detections as a function of a
threshold signal-to-noise ratio $\rho_0$, the Hubble constant $H_0$,
and the binary ``chirp'' mass. For $\rho_0 = 8$, $H_0 = 100$ km/s/Mpc,
and $1.4 \msun$ neutron star binaries, the sample has a median
redshift of $0.22$.  Under the same assumptions but independent of
$H_0$, a conservative rate density of coalescing binaries
($8\times10^{-8}\,{\rm
yr}^{-1}\,{\rm Mpc}^{-3}$) implies LIGO will observe $\sim 50\,{\rm
yr}^{-1}$ binary inspiral events.

The precision with which $H_0$ and the deceleration parameter $q_0$
may be determined depends on the number of observed inspirals.
For fixed mass binary systems,
$\sim 100$ observations with $\rho_0 = 10$ in the
LIGO detector will give $H_0$ to 10\% in an Einstein-DeSitter cosmology,
and 3000 will give $q_0$ to 20\%. For the conservative rate density of
coalescing binaries, 100 detections with $\rho_0 = 10$ will require about
4~yrs.
\end{abstract}


\newacronym{\ligo}{LIGO}{Laser Interferometer Gravitational-wave
Observatory}
\newacronym{\lli}{LLI}{LIGO-like Interferometer}
\newacronym{\ld}{LD}{LIGO detector}

\section{Introduction}
\label{sec:intro}

When completed in the late 1990's, the \ligo\ (Abramovici
{\em et al.\/} 1992) will consist of two interferometers. Concurrently,
the VIRGO consortium (Bradaschia {\em et al.\/} 1990)
will complete a single interferometer of comparable precision.  In
this letter we introduce a new and general class of cosmology tests
based on the anticipated
observation of the gravitational radiation from inspiraling
binary neutron star and/or black hole systems.

Schutz (1986) and Krolak and Schutz (1987)
noted that observation of binary inspiral
in three independent interferometers
will reveal the source's luminosity distance $d_L$.
With an {\em independent\/} measure of the source redshift, such
observations can determine cosmological parameters (e.g., Hubble's
constant $H_0$).  Our tests differ from theirs: the
simplest can determine cosmological parameters with
observations made in a {\em single\/} interferometer and without any
other independent knowledge about individual binary systems.
Extensions systematically exploit additional
information available from multiple
interferometers.

To assess the capabilities of these tests we consider observations in
a single \lli; thus, our work applies directly to the individual
interferometers in the \ligo\ and VIRGO projects. In addition, the two
components making up the \ld\ will be arranged in nearly the same
plane and with nearly identical arm orientations (Whitcomb 1992) -- a
configuration which operates like a single \lli\ of increased
sensitivity (Finn and Chernoff 1993, FC).
We apply our recent results (FC) for the
precision with which a binary's parameters can be
determined in a single
\lli.

Below we describe the physical basis and methodology of
our tests.  As a concrete application we specialize to
neutron star binary systems of constant mass. We find the
effective volumes sampled and the cosmological source rates.
We derive the precision
with which $H_0$ and $q_0$
may be determined in a Friedmann-Robertson-Walker (FRW) universe. We
demonstrate that source evolution may be handled in a straightforward
manner.

\section{Gravitational radiation from inspiraling binaries}
\label{sec:grav-rad}
\typeout{Gravitational radiation from inspiraling binaries}

Assume that quadrupole-formula gravitational radiation adiabatically
damps a binary's Newtonian
orbit.\footnote{Cutler {\em et al.\/} 1992 have
shown that higher order relativistic contributions modify the
orbital evolution. The tests described herein
are entirely compatible with the use of a more
accurate evolution, and we do not expect our
conclusions to be altered significantly by the adoption of a refined
waveform.} Let the distance between the source
and the \lli\ be $d$. The time-dependent
response to the gravity-wave is ($G=c=1$)
\begin{equation}
m(t)={\scrM\over d}\left(\pi\scrM f\right)^{2/3}
\Theta\left(\htns,\htnb\right)
\cos\left( \int^t 2\pi f dt + \Psi\right),
\end{equation}
where the ``chirp'' mass $\scrM = \mu^{3/5}M^{2/5}$ and
$\mu$ and $M$ are the reduced and total mass of the binary system.
The frequency $f = f(\scrM, T-t)$ where
$T$ is the
instant the binary separation $\to 0$ given the above
approximations and $\Psi$ is a
constant phase.
$\Theta$ is a known function of four angles specifying the
orientation of the binary system with respect to the \lli\
($0\leq\Theta\leq4$, FC):
$\htnb$ lies along
the binary's orbital angular momentum and
$\htns$ points from the detector to the binary.
The wave frequency
$f$ is twice the binary's orbital frequency, and a \lli\
will be sensitive to the radiation during the last few minutes of
inspiral as $f$ sweeps from $\sim 10$~Hz to $\sim 1$~KHz.
For binaries at cosmological distances the form of the
response is the same, only now in place of $d$ we have the {\em
luminosity distance\/} $d_L$ and in place of $\scrM$ we have
$\scrM' \equiv (1+z)\scrM$, where $z$ is the redshift.
For more detail, see FC.

Schutz (1986) showed that three interferometers of
different orientations are capable of determining $d_L$,
$\scrM'$, $\Psi$, $T$, $\htns$, and $\htnb$. In contrast, a single
\lli\ can measure only
$\scrA\equiv\Theta/d_L$, $\scrM'$, $T$, and $\Psi$ (FC);
thus, $d_L$ cannot be determined directly since $\htns$ and
$\htnb$ are unknown. However, we know that $\htns$ and $\htnb$ are
each uniformly distributed on the sphere;
consequently, the
distribution of $\Theta( \htns, \htnb )$ is also known.  Thus, the observed
distribution of $\scrA$ is related to the number density of binary
systems on spheres of constant $d_L$.  Similarly, the observed
distribution of $\scrM'$ is connected to the number density on shells
of constant $z$.  With a physically plausible assumption regarding the
evolution of the source distribution function with $z$, the joint
distribution of observed inspiral events as a function of $\scrM'$ and
$\scrA$ is sufficient to measure $H_0$, $q_0$, and otherwise test
cosmological models.

\section{Method}
\label{sec:method}
\typeout{Method}

Let $\Nt d\scrM dV$ be the {\em intrinsic} binary coalescence rate (events
per time), where $\scrM$ is the intrinsic chirp mass and $dV$ is the
local volume element.
The SNR $\rho$ is a function of the parameters
that describe the binary system and its orientation with respect to
the detector.  We give the form explicitly for a single
interferometer in FC (eqs.~3.29, 3.31, and 4.10).
The quantities $dV$ and $\Nt$ depend on the cosmology and the
binary source evolution, and we denote
the full set of model cosmology parameters as $\cosmo$.

The observer counts as detections only those events for which the SNR
exceeds a threshold $\rho_0$,  measuring $\scrM'$, $\rho$, and (for
observations in more than one \lli) additional parameters $\bzeta$ that
are functions of the relative orientations.  We wish to compare the
observed distribution with that implied by a model $\cosmo$.  The
expected event rate density corresponding to the model is
\begin{eqnarray}
\ndot (\scrM', {\hat \rho}, {\bzeta}) & = &
\int d\scrM dV {d\Omega_B\over 4\pi}
{\Nt \over 1 + z} \nonumber \\
& &
\delta[ {\hat \rho} - \rho(d_L, \scrM', \htnb, \htns) ]
\delta[ \scrM' - (1+z) \scrM ]
\delta[ {\bzeta} - {\bzeta}(\htnb, \htns ) ] .
\label{eqn:ndot}
\end{eqnarray}
The integration over $d\Omega_B d\Omega$
effectively averages over
$\htnb$ and $\htns$ since, by assumption, $\Nt$ depends on
neither.
The total rate $\ndot( > \rho_0)$ is
$\ndot (\scrM', {\hat \rho}, {\bzeta})$
integrated over all $\scrM'$, ${\bzeta}$, and $\hat \rho > \rho_0$.
The probability density for observations exceeding the
threshold for a given set of cosmological parameters is
\begin{equation}
p(\scrM', {\hat \rho}, {\bzeta} | > \rho_0, \cosmo ) =
{ \ndot(\scrM', {\hat \rho}, {\bzeta})
\over
\ndot( > \rho_0 ) } .
\label{eqn:prob}
\end{equation}

Given a set of N observations we now apply a standard
maximum likelihood analysis (Eadie {\em et al.\/} 1971). If the
confidence volume $d\scrM'
d\rho dV_\zeta$ is small compared to the scale on
which the probability density $p$ varies, then the total likelihood
$\Lambda(\cosmo) \propto
\prod_{j=1}^N \left( p\, d\scrM'\,d\rho\,dV_\zeta \right)_j $.
{\em The cosmological
parameters $\cosmo$ that best fit the observed distribution are
those that maximize $\Lambda$.\/} Regions with $\Lambda>\Lambda_0$ (for fixed
$\Lambda_0$) are confidence volumes for $\cosmo$.

\section{Application}
\label{sec:application}
\typeout{Application}

As a concrete application, we show how observations in a single
\lli\ can determine the parameters of a matter-dominated FRW model.
Write the volume element $dV$ as $f(z; H_0, q_0) dz d\Omega$,
noting
$f=\tilde{f}(z;q_0)(c/H_0)^3$ and
$d_L=\tilde{d}_L(z;q_0)(c/H_0)$ (where $\tilde{f}$ and
$\tilde{d}_L$ are dimensionless and independent of $H_0$).

Assume that $\Nt$ can be written as
$\Nt = \Nt_0( \scrM ) h(z) (1+z)^3$
where $\Nt_0 (\scrM)$ is the intrinsic rate density for coalescences in
the {\em local} neighborhood.
The factor $h(z) (1+z)^3$ expresses
the time-dependence of the intrinsic rate and
$h(z)$ encompasses all
evolutionary effects for a {\em comoving} volume $[h(0) = 1]$. The
factorization of $\Nt$ implies that the chirp mass
distribution is independent of the age of the universe, although the
coalescence rate may vary.
This assumption is justified if
the character of the binary formation process (suitably averaged over
many galaxies) is independent of $z$.
The assumption is {\em very} plausible
for neutron star binaries.  Neutron stars form in the
collapse of degenerate stellar cores in type II supernovae or by
accretion induced collapse of white dwarfs. In either case the
mechanism that triggers the collapse and determines the mass depends largely
on the equation of state of degenerate matter.
The resultant mass distribution should be relatively insensitive
to the specific evolutionary pathway.
The assumption is {\em moderately}
plausible for binaries containing a black hole. Current understanding
suggests that the hole's upper mass is not tightly constrained
by microscopic physical processes and, in fact, observations show a
range of candidate masses (Casares, Charles and Naylor 1992; Remillard,
McClintock and Bailyn 1992).
The assumption that source evolution is a function of $z$ {\em alone}
is a phenomenological, not a physical, treatment.  In
contrast, a physical approach would introduce additional,
non-cosmological timescales. Then $h$ would depend, not only on $z$, but
also $q_0$ and the timescales for gas cooling, star formation, and so
forth. Here, $h(z)$ serves as a parameterization of the effects of
evolution.

Now assume that neutron star binaries are the predominant
sources. Noting that
the measured range of pulsar masses is small ($\pm 0.1 \msun$; Lamb 1991),
while the derived range of masses in binary X-ray pulsars is somewhat larger
($\pm 0.4 \msun$; Nagase 1989), take
$\Ntz\left(\scrM\right) = \Ntzh\,\delta\left(\scrM-\scrM_0\right)$,
{\em i.e.,\/} assume all binary systems have the same intrinsic chirp mass
$\scrM_0$ (we relax this assumption in our more detailed report).
Since $z$ is a function of
$\chirpM'/\chirpM_0$, we can express $\ndot
(z, \rho )$ as $\ndot(\chirpM', \rho )/\chirpM_0$.
Using FC (eq.~3.16) for $\rho$ in eq.~(\ref{eqn:ndot}), we find
\begin{equation}
\ndot(z | > \rho_0) =
\Ntzh 4 \pi \left( c \over H_0 \right)^3 {\tilde f}(z; q_0) h(z) (1+z)^2
P\left[ \delta { {\tilde d}_L\left( z;q_0 \right ) \over
\left( 1+z \right)^{5/6} } \right] ,
\label{eqn:ndot-z-rho}
\end{equation}
where $P(x)$ is the fraction of randomly oriented binaries
with $\Theta$ exceeding x [tabulated in FC, where it
is called $P(\Theta > x)$]. The distribution is
cutoff as the argument of $P$ increases: $P(0) = 1$
and $P( x > 4)=0$. Note the appearance of
the dimensionless factor
\begin{equation}
\delta = 9.0 \left(\rho_0 \over 8 \right)
        \left(\chirpM_0  \over 1.2 \msun \right)^{-5/6}
           \left(H_0 \over 100\,\km\,\secs^{-1}\,\Mpc^{-1} \right)^{-1}
                     \left( F_{7/3} \over \sqrt{2} \right)^{-1} .
\label{eqn:delta}
\end{equation}
The interferometer's
sensitivity is described by $F_{7/3}$ (for details see FC) which has a
nominal value of $1$ for a single \lli\ of advanced design and, as
used here, $\sqrt{2}$ for the \ld\ composed of two such \lli. We refer to
the parameters in eq.~(\ref{eqn:delta}) as fiducial parameters.
As $\delta
\to \infty$ the fraction of observable binaries goes to zero.  The
total observed rate above threshold is eq.~(\ref{eqn:ndot-z-rho})
integrated over all z.

Assuming no source evolution [$h(z) = 1$], expand $\ndot(z | > \rho_0)$
to lowest order in $z$ to obtain the  Euclidean rate
\begin{equation}
\ndot_E( > \rho_0) = \Ntzh 4 \pi \left( c \over H_0 \right)^3
\left( 1.836 \over \delta^3 \right) ,
\label{eqn:ndot-tot-euclidean}
\end{equation}
independent of $H_0$ and
$\propto 1/\rho^3$ above threshold. We summarize observation rates in
terms of $Q(q_0, \delta, h) = \ndot(> \rho_0) / \ndot_E( > \rho_0)$.
Note $\lim_{\delta \to \infty} Q = 1$.

\section{Results}
\label{sec:results}
\typeout{Results}

Consider first the characteristic depth of the observed sample.
Let $z_{1/2}$ ($z_{max}$) be the redshift at which $P = 0.5$ ($0.0$)
in eq.~(\ref{eqn:ndot-z-rho}). To lowest order in redshift,
$z_{1/2} = 0.145 (9.0/\delta)$ and $z_{max} = 0.44 (9.0/\delta)$.
Note that we have scaled $\delta$ to the fiducial parameters of eq.
(\ref{eqn:delta}). The observed sample is more precisely characterized
by the cumulative distribution of {\em detected} binaries and by its
{\em completeness}, displayed in figure 1a for $\delta = 4.5$, $9.0$,
$13.5$ and $18.0$ (SNR thresholds of $4.0$, $8.0$, $12.0$ and $16.0$
for fiducial parameters). Koo and Kron (1992, KK) model the
distribution of faint galaxies without source evolution but with a low
$H_0$.  To compare with KK adopt $H_0 = 50$ km/s/Mpc; then the median
observed binary has $z = 0.10$ at $\rho_0 = 8$ ($\delta = 18.0$).  In
KK, the median of the cumulative galaxy distribution with $18 < B < 19$
occurs at roughly the same redshift.  The depth is modest, but (as
figure 1a shows) the extent is large.  The cumulative distribution for
different values of $q_0$ show discernible changes.

The detection rate $\ndot( > \rho_0)$ scales directly with the local
coalescence rate.  Phinney (1991) has discussed the observational and
theoretical limits.  Based on the several observed neutron star
binaries, the time until coalescence, and the effective Galactic
volumes sampled, he gives a ``best guess'' estimate for the coalescence
rate density of $8 \times 10^{-8}$ Mpc$^{-3}$ yr$^{-1}$. This is a {\em
conservative} estimate because it counts only binaries like those
observed to date; for a number of reasons the true value may exceed the
estimate by as much as $\sim 800$ [$H_0/$($100$ km/s/Mpc)]$^3$.
Scaling to the conservative value, the anticipated detection rate is
\begin{equation}
\ndot( > \rho_0) = 68. { \Ntzh \over 8 \times 10^{-8} \Mpc^{-3} \secs^{-1} }
			Q( q_0, \delta, h)
                 \left( 8 \over \rho_0 \right)^3
                 \left( \chirpM_0 \over 1.2 \msun \right)^{5/2}
		 \left( F_{7/3} \over \sqrt{2} \right)^{3} .
\end{equation}
Recall that the function $Q$ is the ratio of the cosmological rate to
the Euclidean rate; figure 1b gives $Q$ for a range of $\delta$ for
values of $q_0 = 0.25$, $0.50$, and $0.75$ assuming no source evolution.
For the fiducial parameters, $Q = 0.715$ for $q_0 = 0.5$, which
corresponds to an observed rate of $\ndot(>8) = 49.$ yr$^{-1}$ events. Notice
$Q$ is only weakly sensitive to $q_0$: in an analytic expansion to
lowest order in $z$ it is independent of $q_0$. The effects of
source evolution may dominate the influence of $q_0$ on the value of
$Q$.

Our main interest here is to assess our
cosmology test when source evolution takes place; consequently we
adopt an {\em ad hoc} model $h(z) = \exp[ -( \alpha z + \beta z^2) ] $ for
$z < 10$ and $h(z)=0$ otherwise. This model
is useful for both strong and weak evolution.
Broadhurst {\em et al.\/} (1988) consider a similar form for galaxy
luminosity evolution; however, we make no attempt to apply their
results. In adopting this model, we envision three qualitatively
different possibilities: (1) The coalescence rate may be simply
proportional to the star formation rate. Then, based on the faint
galaxy studies, there should be little evolution for $z < 0.5$ (KK).
We assume $\alpha = \beta = 0$ for {\em uniform production} (no
evolution).  (2) Neutron star binaries may be formed by collisional
encounters in the dense cores of galaxies.  The luminosity evolution of
QSOs is a strong function of $z$ (Boyle {\em et al.\/} 1987, 1988);
hence, if the coalescence rate is proportional to the QSO luminosity, a
substantial increase in the production rate with $z$ is expected.  We
take $\alpha = -3/2$ and $\beta = 1/4$ [for which $h(z)$ peaks at $z =
3$, where it is $h \approx 9.5$] for this case of {\em early
production}.  (3) Binaries may be formed by collisional encounters in
globular clusters after core collapse.  The time to core collapse is of
order $1/H_0$ for clusters in the Galaxy. For the case of {\em late
production}, we take $\alpha = 0$ and $\beta = 1$, which cuts off
strongly for $z > 1$.  Figure 1b illustrates how early and late
production alters $Q$ in the $q_0 = 0.5$ model.

We used a Monte Carlo analysis to calculate the precision with which $
\cosmo $ can be determined.  We chose ``true'' model parameters $
\widetilde{\cosmo} $ and formed a realization of N observations.  Since
we have restricted our analysis to large $\rho_0$ the small
observational uncertainties have not been included in the
realization.\footnote{FC show that the fractional errors
$d\chirpM'/\chirpM'\approx 10^{-4}/\rho$ and $d\rho/\rho \approx
1/\rho$.} We applied a maximum likelihood procedure to this realization
to deduce the most likely values $ \widehat{\cosmo} $, and found the
difference $\delta{\cosmo} = {\widehat{\cosmo}} - {\widetilde{\cosmo}}$
as a function of $N$ by repeating the procedure 100 times for each
$N$.  For each parameter $\mu_i$, we found the bias $\left<\delcosmo
\right>$ and the rms dispersion $\sigma_{\mu_i} \equiv \left<
\delcosmo^2 \right>^{1/2}$. In all cases we found the magnitude of the
bias to be much less than the dispersion; so, we deal with the latter
as an indication of the best performance that can be expected for
measurement of $\cosmo$.

Our model depends upon five physical quantities: Hubble's constant,
the deceleration parameter, the chirp mass, $\alpha$ and $\beta$.
Denote the ``true'' values by $\{ {\widetilde H_0}, {\widetilde q_0},
{\widetilde \chirpM_0}, {\widetilde \alpha}, {\widetilde \beta} \}$
and write the cosmological parameters as $\{ \cosmo \} = \{ {h_0
\equiv H_0/\widetilde H_0}, q_0, m \equiv {\chirpM/ \widetilde\chirpM_0},
\alpha, \beta \}$. (In this dimensionless form $\{ \widetilde{\cosmo} \}
= \{ 1, {\widetilde q_0}, 1, {\widetilde \alpha}, {\widetilde
\beta}\}$.)  We carried out the Monte Carlo calculations of the rms
dispersion of $\widehat{h_0}$ and $\widehat{q_0}$ for
Einstein-DeSitter cosmologies ({\em i.e.}, ${\widetilde q_0} = 0.5$).
Our results are unchanged by scaling $\widetilde H_0$, $\widetilde
\chirpM_0$, $\rho_0$ and $F_{7/3}$ for fixed $\delta$. We assumed uniform {\em
a priori} probabilities for $h_0\in (0.1,2.0)$ and $q_0\in(0.1,2.0)$;
the allowed ranges of $m$, $\alpha$, and $\beta$ were large and will
be discussed in more detail elsewhere.

Table~\ref{tbl:res} summarizes results for
$\sigma_{h_0}$ and $\sigma_{q_0}$ for two
values of $\delta$, three choices of $\alpha$ and $\beta$, and several
different $N$.  While the dispersion and bias in $h_0$, $q_0$,
$\alpha$ and $\beta$ are all strongly correlated, we defer a more
detailed discussion to a forthcoming paper.  The dispersions reported
in Table 1 should only be taken as rough approximations: we have not
attempted to determine the precision to which they are found by our
Monte Carlo analysis. It is clear, however, that our samples (for
fixed $\delta$) extend into the asymptotic regime, where the
dispersion is $\propto N^{-1/2}$.

Note how, at fixed $N$,
the Hubble constant is not better determined by the deeper sample
(smaller threshold $\rho_0$, which implies smaller $\delta$): thus,
sample size is more important than sample depth to the accurate
determination of $h_0$. In contrast, the uncertainties in
$q_0$ are significantly smaller in the deeper sample. More quantitatively,
we find that with a threshold $\rho_0 \sim 10.7$ ($\delta = 12$),
LIGO will determine $H_0$ to $\sim 10$\% with
$\sim 100$ observations. Similarly, $q_0$ may be determined to approximately
$\sim 20$\% with $\sim 3000$ observations.

\section{Discussion and conclusions}

Observations of the gravitational radiation from binary inspiral by
the LIGO detector can be used to measure important cosmological
parameters. These measurements rely on comparing the observed
distribution of inspiral events, as a function of signal strength and
chirp mass, with model predictions. It is necessary to
adopt a model for the intrinsic coalescence rate of binaries as a
function of the chirp mass and the redshift. We argue that a
factorized form is plausible but that other forms are possible. We
provide a simple demonstration of the power of the test by
specializing to the case of a narrow, well-defined chirp mass range.
We characterize the observational volumes by the sample's median
redshift and by its completeness. We
give the rate of detections as a function of the detector
threshold signal-to-noise ratio using Phinney's (1991) conservative
estimate of the local binary coalescence rate.  We find that, with
a modest number of detections, these observations can determine
the Hubble constant to 10\%. The conservative rate estimate implies
that the observations would take $\sim 4$ years.  We
address the optimal choice of the SNR threshold,
measurement errors at
small $\rho$ and a range of binary
masses in a paper in preparation.

\acknowledgements

We thank Drs.~Bildsten, Cutler, Flanagan, Loredo, Thorne, Wasserman,
and Whitcomb for discussions.  LSF acknowledges the Sloan
Foundation and NASA NAGW-2936 (Northwestern); DFC the PYI program,
NSF AST-8657467 and NASA NAGW-2224 (Cornell).


\begin{references}
\reference
Abramovici, A. et al., 1992, {\em Science}, {\bf 256}, 325.

\reference
Bradaschia, C. et al., 1990, {\em Nucl. Inst. A.}, {\bf 289}, 518.

\reference
Broadhurst, T. J., Ellis, R. S., Shanks, T. 1988, {\em M.N.R.A.S.}, {\bf 235},
827.

\reference
Boyle, B. J., Fong, R., Shanks, T. and Peterson, B. A. 1987, {\em M.N.R.A.S.},
{\bf 227}, 717.

\reference
Boyle, B. J., Shanks, T. and Peterson, B. A. 1988, {\em M.N.R.A.S.},
{\bf 235}, 935.

\reference
Casares, J., Charles, P. A., and Naylor, T. 1992, {\em Nature}, {\bf 355}, 614.

\reference
Cutler, C., Finn, L. S., Poisson, E. and Sussman, G. J. 1993,
{\em Phys.\ Rev.\ D.,} {\bf 47,} 1511--1518.

\reference
Eadie, W. T., Drijard, D., James, F. E., Roos, M. and Sadoulet, B. 1971,
{\em Statistical Methods in Experimental Physics}, North Holland, 121.

\reference
Finn, L. S. and Chernoff, D. F. 1993, {\em Phys.\ Rev.\ D.,\/}
{\bf 47,} 2198--2219.

\reference
Koo, D. C. and Kron, R. G. 1992, {\em Ann. Rev. Astron. Astophys.},
{\bf 30}, 613.

\reference
Krolak, A. and Schutz, B. F. 1987, {\em Gen. Rel. and Grav.}, {\bf 19}, 1163.

\reference
Lamb, F. K. 1991, Neutron Stars and Black Holes in {\em Frontiers of
Stellar Structure}, ed. D. L. Lambert (ASP), 299.

\reference
Nagase, F. 1989, {\em Publ. Astr. Soc. Japan} {\bf 41}, 1.

\reference
Phinney, E. S. 1991 {\em Astrophy. J.}, {\bf 380}, L17.

\reference
Remillard, R. A., McClintock, J. E., and Bailyn, C. D. 1992, {\em
Astrophy. J.}, {\bf 399}, L145.

\reference
Schutz, B. F. 1986 {\em Nature}, {\bf 323}, 310.

\reference
Whitcomb, S. 1992, private communication.

\end{references}

\eject
{\centerline{FIGURES}}

Figure 1a. The observational
volume accessible to the \ld\ is characterized by the cumulative
distribution of {\em detected} binaries $S(z,\rho_0) = \int_0^z dz'
\ndot( z' | > \rho_0 ) / \ndot( > \rho_0 )$ (the rising set of curves)
and by the sample's {\em completeness} $ T(z,\rho_0) =
\int_0^z dz' \ndot( z' | > \rho_0 ) / \int_0^z dz' \ndot( z' | > 0 ) $
(the falling curves) for the case of no source evolution. The widely
spaced families of curves represent $\delta = 4.5$, $9.0$, $13.5$ and
$18.0$ (eq. \ref{eqn:delta}) from right to left; these correspond to
thresholds of $\rho_0 = 4$, $8$, $12$ and $16$ for fiducial
parameters. Each family consists of three closely spaced curves
corresponding to different values of $q_0$ (dashed $0.25$; solid
$0.5$; dotted $0.75$). The curves describing the sample's completeness
terminate at the maximum $z$ at which an inspiraling binary can be
observed.

Figure 1b. The ratio of the cosmological to the Euclidean rate
of detections, $Q$, is given as a function of $\delta$ (eq.
\ref{eqn:delta}).  For the case of no source evolution, three
different values of $q_0$ are displayed, as in
Figure 1a; these correspond to the three closely spaced lines. Two
additional models discussed in the text are shown for the $q_0 = 0.5$
case.  The long dashes illustrate a coalescence rate lower
in the past than in the present. The chain dashes illustrate one with
a higher rate in the past.

\eject
\begin{table}
\caption{The results of applying the maximum-likelihood analysis to
finding $h_0$, $q_0$, $m_0$, $\alpha$ and $\beta$. }
\label{tbl:res}
\begin{tabular}{cccccccc}
$\delta$ & $N$ & \multicolumn{2}{c}{No Evolution}
               & \multicolumn{2}{c}{Early Production}
               & \multicolumn{2}{c}{Late Production}\\
\multicolumn{2}{c}{} &
\multicolumn{2}{c}{($\alpha=\beta=0$)}&
\multicolumn{2}{c}{($\alpha=0$, $\beta = 1$)}&
\multicolumn{2}{c}{($\alpha=-1.5$, $\beta=0.25$)}\\
\multicolumn{2}{c}{} &
$\sigma_{h_0}$&
$\sigma_{q_0}$&
$\sigma_{h_0}$&
$\sigma_{q_0}$&
$\sigma_{h_0}$&
$\sigma_{q_0}$\\
\tableline\\
 12.0&   30  &0.144 &0.844 &0.143 &0.838          &0.130 &0.881 \\
     &  100  &0.099 &0.681 &0.093 &0.674	   &0.104 &0.666 \\
     &  300  &0.071 &0.479 &0.063 &0.425	   &0.075 &0.425 \\
     & 1000  &0.035 &0.197 &0.039 &0.226	   &0.036 &0.193 \\
     & 3000  &0.019 &0.110 &0.017 &0.102	   &0.020 &0.110 \\
     &10000  & -    & -    &0.009 &0.055	   &0.010 &0.054 \\
 20.0&   30  &0.097 &0.890 &0.112 &0.865          &0.109 &0.896 \\
     &  100  &0.075 &0.780 &0.076 &0.742	   &0.067 &0.696 \\
     &  300  &0.054 &0.590 &0.049 &0.515	   &0.054 &0.558 \\
     & 1000  &0.029 &0.311 &0.026 &0.255	   &0.027 &0.273 \\
     & 3000  &0.013 &0.141 &0.015 &0.157          & -    & -    \\
\tableline
\end{tabular}
\end{table}
\end{document}